\documentclass[aps,prl,twocolumn]{revtex4-1}
\usepackage{graphicx,epsf, epsfig, amssymb, amsmath}
\usepackage{ulem,color}
\usepackage{chapterbib}

\newcommand{\aap}{Astron.\ Astrophys.}
\newcommand{\mnras}{Mon.\ Not.\ R.\ Astron.\ Soc.}

\newcommand{\nphysa}{Nucl.\ Phys.\ A}
\newcommand{\apss}{Astrophys.\ Space Sci.}
\newcommand{\sovast}{Sov.\ Astron.}

\newcommand{\acdel}[1]{} 

\begin{document}

\title{Thermodynamically consistent equation of state for an accreted neutron star crust}

\author{M.E.\ Gusakov}
\affiliation{Ioffe Institute, St.-Petersburg, Russia}

\author{A.I.\ Chugunov}
\affiliation{Ioffe Institute, St.-Petersburg, Russia}

\date{\today}

\begin{abstract}
We study equation of state (EOS) of an accreting neutron star 
crust.
Usually, such EOS is obtained assuming (implicitly)
that the free (unbound) neutrons and nuclei in the inner crust  move together.
We argue, that 
this assumption violates the condition $\mu_n^\infty={\rm const}$,
required for hydrostatic 
(and diffusion) equilibrium 
of unbound neutrons
($\mu^\infty_n$ is the redshifted neutron chemical potential).
We construct a new EOS respecting this condition,
working in the compressible liquid-drop approximation.
We demonstrate that it is close to the catalyzed EOS in most
part of the inner crust, being very different from EOSs of accreted crust 
discussed in the literature.
In particular, 
the pressure at the outer-inner crust interface does not 
coincide with the neutron drip pressure, usually calculated in the literature,
and is determined 
by hydrostatic (and diffusion) equilibrium conditions within the star.
We also find an instability at the bottom of fully accreted crust
that transforms nuclei into homogeneous nuclear matter.
It guarantees that the structure of fully accreted crust remains self-similar 
during accretion.
\end{abstract}


\maketitle

\textit{Introduction.--}
%
Neutron stars (NSs) are the densest objects in the Universe. 
The composition of their deepest layers (inner core)
is uncertain and is considered as
the main mystery of the NS physics \cite{hpy07}.
In contrast, it is believed that the composition of their outer layers,
the so called NS crust,
is relatively well-known.
The outer region of the neutron star crust, referred to as the outer crust, 
is composed of atomic nuclei on the neutralizing background of degenerate, almost ideal electron gas. 
In the deeper layers, called the inner crust,
the unbound neutrons
are also present \cite{hpy07,ch08}.
With the subsequent increase of the density, 
the crust ends and we reach the outer core, 
which (in the vicinity of the crust-core boundary) is composed of 
neutrons ($n$), protons ($p$), and electrons ($e$).

The only way to check whether theoretical models of superdense matter are reliable 
is to confront them with  
NS observations.
One of the most promising possibilities 
in this regard 
is to study accreting NSs, 
which 
are observed
in binary systems with the active mass transfer from a companion star.
For some of these sources the accretion process is transient, 
and in quiescent periods X-ray telescopes are able to detect 
the thermal emission from the NS surface, revealing that it is heated up by accretion \cite{heinke_et_al_09,wdp17,Zhao_ea19_DeepCrustalHeatingInHMXB,pcc19}. 
It is generally believed, that the heating is caused by the non-equilibrium nuclear reactions, 
which are initiated in the crust 
as 
it is compressed 
under the weight of newly accreted material.
Obviously, adequate interpretation of observations requires a reliable model describing this process 
(see \cite{mdkse18} for a recent review). 
A similar process can also be important 
for reheating of millisecond pulsars \cite{gkr15}.

Starting with Ref.\ \cite{Sato79}, 
a number of authors study the evolution 
of an 
accreted element
as it compresses and sinks 
deeper and deeper towards the NS core
in the course of accretion.
Some of them used a one-component approximation  
\cite{HZ90,HZ90b,HZ03,HZ08,Fantina_ea18}, 
while others used reaction networks, 
allowing for mixtures of different nuclei \cite{Steiner12,lau_ea18,SC19_MNRAS,SC19_JPCS}. 
They applied either liquid-drop models \cite{HZ90,HZ90b,HZ03,HZ08,SC19_JPCS,Steiner12}, or
up-to-date theoretical atomic mass tables \cite{lau_ea18,SC19_MNRAS}, 
or detailed extended Thomas-Fermi calculations, 
allowing for the existence of unbound neutrons \cite{Fantina_ea18}.
The main common feature of   
all these works is that 
they follow compositional changes associated with reactions induced by the increasing pressure inside an accreted fluid element (`traditional approach').
Such consideration would be clearly applicable 
if we compress uniform infinite matter.
But in reality
the inner crust is not uniform, 
so that
unbound neutrons can 
travel
between different layers to lower the system energy
\footnote{The only exception, which allows for diffusion of unbound neutrons is 
a series of works
\cite{BKKC76,BKC79}, 
but it mainly focuses on the crust properties of newly born NSs
(see also Discussion section in Ref.\ \cite{Steiner12})}.

The traditional approach was known to lead to jumps of 
the neutron chemical potential $\mu_n$ at the phase transitions, 
which are especially 
pronounced
in the one-component approximation 
and 
considerably soften
if 
mixtures of nuclei
are allowed for \cite{Steiner12}.
However, 
these effects were typically 
considered as
a local inconsistency, 
which, likely, does not affect the global properties of the accreted crust. 
In this Letter we show that it is not the case 
and that allowing unbound neutrons
to move independently of nuclei
has a dramatic
effect on the crust composition and equation of state (EOS).
We construct the corresponding EOS
within the 
compressible liquid drop model (CLDM), which 
ignores pairing and shell effects.
This EOS is fully thermodynamically consistent,
in particular, 
$\mu_n$ in the inner crust is continuous and, moreover, 
satisfies the hydrostatic (and diffusion) equilibrium condition, 
$\mu_n^\infty={\rm const}$ (see below),
where 
$\mu_n^\infty \equiv \mu_n \, {\rm exp(\nu/2)}$ is the redshifted $\mu_n$,
and 
$\nu=2\phi/c^2$ ($\phi$ is the gravitational potential and $c$ is speed of light \cite{hpy07}).

The calculated EOS 
significantly differs from EOSs obtained 
within the traditional approach, being very close to the EOS of catalyzed crust.

In Ref.\ \cite{CS19_NoEquil}
we demonstrate additional inconsistency of the traditional approach:
It leads to strong violation of the force balance equation for nuclei 
(gravitational and electric forces are both directed downwards)
in a few rather extended regions of the inner crust, thus
revealing inconsistency of the traditional approach from another point of view.

\textit{Equilibrium condition for unbound neutrons. --}
Neutrons, not bound to nuclei, exist in the inner NS crust.
Except for a narrow layer of width $L$ near the outer-inner crust interface 
($L\lesssim 5$~m for $T=5\times10^8$~K), 
they are superfluid (e.g., Ref.\ \cite{ch08}), and move
with the velocity $V_{sn}$,
governed by the (linearized) superfluid equation,
$m_n \partial V_{sn}/\partial{t}=-\nabla \mu_n^\infty$,
where $m_n$ is the neutron mass
(see, e.g., Refs.\ \cite{khalatnikov89,ga06,pcr10}).
The hydrostatic equilibrium implies $\mu_n^\infty=\mathrm{const}$ 
as a necessary condition in the whole region of neutron superfluidity.

In the narrow nonsuperfluid layer the typical diffusion time, 
$\tau^D\sim L^2/D\lesssim 3\times 10^6$~s 
(the neutron diffusion coefficient $D$ is  
estimated in analogy to Ref.\ \cite{BKC79}), is
much smaller 
than the 
replacement timescale of this layer by accretion, 
$\sim \rho L/\dot M \gtrsim 2\times 10^{9}$~s
(we take $\rho\sim 4.3\times 10^{11}$~g~cm$^{-3}$ for the density 
and assume that $\dot{M}$ 
equals the local Eddington accretion rate, $\dot M\sim 10^5$~g\,cm$^{-2}$\,s$^{-1}$).
As a result, 
unbound neutrons in the non-superfluid layer should be, to a good approximation, 
in diffusion equilibrium: $\mu_n^\infty=\mathrm{const}$.

\textit{Three crustal EOSs .--}
%
Typical temperature in the crust of accreting NSs is $T\lesssim 5\times 10^8$~K
and have a minor effect on EOS \cite{hpy07},
so below we shall work in the approximation of $T=0$.
As discussed above, 
neutron Hydrostatic and Diffusion (nHD) equilibrium conditions imply  
$\mu_n^\infty=\mathrm{const}$ in the whole inner crust
(we assume that the region of neutron superfluidity extends to the crust-core boundary). 
To illustrate importance of this condition 
let us consider three EOSs: 
catalyzed (ground state) EOS,
which is believed to describe 
pristine NS crust
and two EOSs for accreted NS crust: 
(i) traditional, which completely disregards neutron diffusion 
(denoted as `Trad' EOS)  
and (ii) new EOS that respects the nHD condition (denoted as `nHD' EOS).
For simplicity we apply 
CLDM, in which  nuclei are described as liquid drops, 
located at the center of the spherical Wigner-Seitz (WS) cells  \cite{lpr85,hpy07,ch08}.  
We ignore a possible layer of nonspherical nuclei 
in the vicinity of the crust-core boundary 
(for EOSs based on SLy4 energy density functional, 
employed here, this layer is absent \cite{DH00,Vinas_ea17}).
We also assume that the proton drip does not occur in the crust, 
which is true for all numerical models discussed here.
The model is parametrized by
the number densities $n_{ni}$, $n_{pi}$, and $n_{no}$ for, respectively,
neutrons and protons inside, and neutrons outside nuclei;
the neutron skin surface density $\nu_{s}$;
and the volume $V_c$ of WS cell, as well as by the (proton) radius $r_p$ of a nucleus.
In addition, it is useful to introduce the volume fraction 
occupied by nucleus inside the WS cell, $w=4 \pi r_p^3/(3V_{c})$; 
the surface area of a nucleus, $\mathcal A=4\pi r_p^2$; 
and the electron number density $n_e$, 
determined from the quasi-neutrality condition, $n_e=w n_{pi}$.
Within CLDM the energy density can be written as
\begin{eqnarray}
\epsilon&=&w\, \epsilon^{\rm bulk}(n_{ni},\, n_{pi})+(1-w)\,
\epsilon^{\rm bulk}(n_{no},\, 0)
\nonumber\\
&&+E_{s}(\nu_{s},r_{p})/V_{c}
+E_{C}(n_{pi},r_{p},w)/V_{c}
+\epsilon_e(n_e).
\label{Energy}
\end{eqnarray}
Here $\epsilon^{\rm bulk}(n_n,n_p)$ is the energy density of homogeneous nuclear matter;
$E_{s}$ is the surface energy of a nucleus.
The Coulomb energy of a WS cell is given by
$E_C=(16\,\pi^2/15)(n_{pi} e)^2 r_p^5 f(w)$,
where $f(w)=1-1.5\,w^{1/3}+0.5\ w$, 
and $\epsilon_e$ is the energy density of degenerate electron gas  \cite{hpy07}.

Taking the baryon number density,
$n_b=w\,(n_{pi}+n_{ni})+(1-w)\,n_{no}+\mathcal A \nu_s/V_c$, 
and number density of nuclei, $n_N=V_c^{-1}$, to be fixed, 
we minimize $\varepsilon$ with respect to other independent variables
and obtain the beta-equilibrium, mechanical and local neutron diffusion equilibrium (within one unit cell) conditions.
Using these conditions
(see Supplementary material for more details),
we arrive at the two-parameter equation of state, $\epsilon=\epsilon(n_b, \, n_N)$, 
with the second law of thermodynamics 
presented as
\begin{equation}
\mathrm d \epsilon =\mu_n \mathrm{d} n_b +\mu_N \mathrm d n_N,
\end{equation}
where $\partial \epsilon(n_b,\,n_N) /\partial n_b$ is denoted as $\mu_n$, 
because it
equals the chemical potential of 
free (unbound) neutrons, 
as follows from the
minimization procedure discussed above.
The effective chemical potential $\mu_N$ describes 
the energy change due to addition of an extra nuclear cluster to the system at fixed $n_b$,
$\mu_N=(\sigma \mathcal A- 2 E_C)/3$, 
where $\sigma$ is the surface tension \cite{hpy07} 
(see Supplementary material).
Catalyzed EOS corresponds to absolute minimum of $\varepsilon$ at fixed $n_b$, 
hence it is given by the condition $\mu_N=0$. 
With this condition EOS becomes one-parametric, i.e.,
specified in a unique way for a given $n_b$. 
 
For accreted crust $T$ is not high enough 
to allow for nuclear reactions that minimize $\varepsilon$
by choosing 
$n_N$ in an optimum way,
thus $\mu_N$ is, generally, non-zero.
To make EOS one-parametric 
we need an additional 
equation.
In the traditional approach (e.g., Ref.\ \cite{HZ90})
the equation
follows from the requirement that 
the total baryon number in the WS cell is conserved, 
$A_c=n_b V_c=\mathrm{const}$.
(Note that this 
equation
should be modified in the regions where pycnonuclear reactions
proceed and $A_c$ doubles \cite{HZ90,HZ03,HZ08,Fantina_ea18}.)

And what about nHD EOS?
$A_c$ is not conserved now, because neutrons can move independently of nuclei.
Instead, this EOS should respect the nHD condition $\mu_n^\infty={\rm const}$,
as well as the general hydrostatic equilibrium condition, 
$P^\prime(r)=-(P+\epsilon) \nu^\prime(r)/2$
 \cite{hpy07},
where $P$ is the pressure and prime ($^\prime$) means derivative with respect to the radial coordinate $r$.
Combining these two equations with the Gibbs-Duhem relation, $dP= n_b\, d\mu_n+n_N\, d\mu_N$,
one arrives at 
the requirement 
$\mu_\mathrm{N}^\infty=\mu_\mathrm{N} e^{\nu/2}={\rm const}$.
In other words (because 
$\mu_n e^{\nu/2}$
is also a constant),
the ratio $\mu_\mathrm{N}/\mu_n$ must be fixed in the inner crust, i.e., 
$\mu_\mathrm{N}/\mu_n=C$, or, recalling the definition of $\mu_N$,
\begin{equation}
\sigma \mathcal A- 2 E_C= 3\, C \mu_n,
\label{diff}
\end{equation}
where $C$ is some constant.
This condition 
parametrizes a family of nHD EOSs.
It allows, in  particular, to present $\mu_n$ 
as a function of $P$ and $C$: $\mu_n=\mu_n(P,\, C)$.
Catalyzed 
EOS is a member of this family
(hence neutrons are in the diffusion equilibrium in catalyzed matter -- an expected result);
it corresponds to the choice $C=0$ (i.e., $\mu_\mathrm{N}=0$). 
As shown below, only one 
particular $C$ corresponds to  
the fully accreted NS crust, 
which we shall be mostly interested in what follows. 

\textit{nHD EOS for a fully accreted crust. --}
%
In this case $C$
can be determined 
from two requirements: (i) $P$ and $\mu_n$ at the crust-core boundary must be continuous;
and (ii) the structure and composition of fully accreted crust 
should not change
in the course of accretion.
In particular, the latter condition means that the total number of nuclei in the crust should be conserved. 
However, accretion permanently 
brings
nuclei to the crust. 
Clearly, the stationary situation is possible only
if the same amount of nuclei disintegrate 
somewhere in the crust.

nHD EOS provides a natural mechanism of nuclei disintegration 
due to a specific instability
discussed below.
Namely, 
numerical calculations show
that, 
for each $C$
there is a maximum pressure 
$P_{\rm max}$,
such that the solution to Eq.\ (\ref{diff}) does not exist at $P>P_{\rm max}$
(in Supplementary material we argue that it is a general feature of nHD EOSs).

To demonstrate the physical mechanism behind the instability 
we, 
first of all, combine 
the equation
$P^\prime(r)=-(P+\epsilon) \nu^\prime(r)/2$
and condition $\mu_n^\infty={\rm const}$
to derive a relation, $d\, \mu_n=\mu_n/(P+\epsilon) d\, P$, 
which is valid in the 
nHD-equilibrated
inner crust and is {\it equivalent} to Eq.\ (\ref{diff}). 
It states that $\mu_n$ in a given volume is fixed if $P$ is fixed,
independently of nuclear transformations occurring in this volume.  
Now, let us consider a layer, initially located at $P_\mathrm{max}$, 
but compressed slightly by newly accreted material, 
so that $P$ is a bit larger than $P_{\rm max}$.
Absence of stationary solutions at $P>P_\mathrm{max}$ 
means that the 
layer
should be out of beta-equilibrium 
at such pressure 
(otherwise it is impossible to remain in the
hydrostatic equilibrium).
Then beta-captures 
come into play trying to return the system to the beta-equilibrium, but, as we checked numerically, they are accompanied by neutron emissions and the emitted neutrons	diffuse out of the layer in order to preserve $\mu_n$ at a given $P$.
As a result, the layer begins to shrink 
and nuclei in the layer start to `evaporate' 
($A$ and $Z$ decrease)
until disintegration -- the required instability.

This instability is, in fact, 
similar to the mechanism 
discussed
in Ref.\ \cite{BKC74}. 
Namely, at $P>P_\mathrm{max}$ nuclei 
become
unstable 
with respect to electron capture accompanied by emission of neutrons. 
Each electron capture makes nucleus even more unstable, 
leading to a series of subsequent electron captures and neutron emissions 
until complete disintegration. 
The instability is also analogous to the 
superthreshold electron capture cascades
studied in Refs.\ \cite{Gupta_ea07,lau_ea18,SC19_MNRAS}, 
but, in contrast to these works, 
disintegration is 
complete and takes place at fixed $P$~and~$\mu_n$.

During accretion, the number of nuclei in the (initially catalyzed) crust 
is increasing until the instability sets in
at $P=P_{\rm max}$.
Since at $P>P_{\rm max}$ stable crust does not exist,
$P_{\rm max}$ should coincide with the pressure at the crust-core boundary
\footnote{This is strictly true for CLDM employed here. 
For a more realistic models with integer $A$ and $Z$ and, possibly, 
with 
nonspherical nuclei
\cite{rpw83_Pasta} near the crust-core boundary, 
it may happen that the instability occurs earlier, 
e.g., at the phase transition between 
the (standard) spherical nuclei and cylindrical nuclear shapes.}.
Thus, 
the parameter $C$
and hence nHD EOS for a fully accreted crust 
(hereafter, simply `nHD EOS')
can be determined 
by matching $\mu_n$ at $P=P_{\rm max}$
in the crust and in the core:
$\mu_n(P_{\rm max},\, C)=\mu_n^{\rm core}(P_{\rm max})$,
where $\mu^\mathrm{core}_n(P)$ 
stands for $\mu_n$ in the core.

Now we have everything at hand 
to find where 
the outer-inner crust interface is located.
To this end, we note that, by definition,
one has $m_n=\mu_n$ at the interface,
thus the pressure $P_{\rm oi}$ there 
can be found from the condition: $m_n=\mu_n(P_\mathrm{oi},\, C)$.
Note that it should not necessarily coincide (for nHD EOS) 
with the neutron drip pressure $P_{\rm nd}$
of Trad EOS,
because the latter is obtained neglecting 
possible redistribution of neutrons in the star.

\begin{figure}
	\includegraphics[width=\columnwidth]{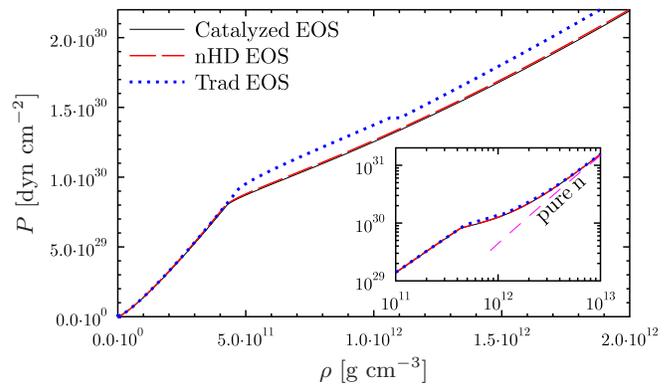}%
	\caption{Pressure versus density 
		 for different crustal EOSs discussed in the text. \label{Fig_EOS}}
\end{figure}

\textit{Numerical example. --}
%
To illustrate our results 
we employ SLY4 energy density functional \cite{Chabanat_ea98_SLY4};
the corresponding surface energy and tension $\sigma$ are adopted from Ref.\ \cite{DHM00}.
We find that the interface between the inner and outer crust is located at 
$8.0\times 10^{29}$,  $8.1\times 10^{29}$, and $9.1\times 10^{29}$~dyn\,cm$^{-2}$ for catalyzed, nHD, and Trad EOSs, respectively
(for nHD EOS such $P_{\rm oi}$ leads to $C\approx 0.0025$).
The corresponding pressures at the crust-core boundary 
equal $4.93\times 10^{32}$, $5.20\times 10^{32}$, and
$5.14\times 10^{32}$~dyn\,cm$^{-2}$.
For simplicity, when considering  Trad model, 
we assume that the pycnonuclear reactions take place at $Z=10$.

Fig.\ \ref{Fig_EOS} demonstrates three EOSs described in this work: 
catalyzed (solid line), nHD (long dashes), and Trad (dots);
EOS of pure neutron matter (dashed line) is added for comparison. 
One can see that nHD EOS, suggested here, significantly differs from Trad EOS, 
obtained within the traditional approach, and is much closer to the catalyzed EOS
(cf., e.g., $P_\mathrm{oi}$ for catalyzed and nHD EOSs: 
$8.0\times 10^{29}$ and $8.1\times 10^{29}$~dyn\,cm$^{-2}$).

\begin{figure}
	\includegraphics[width=\columnwidth]{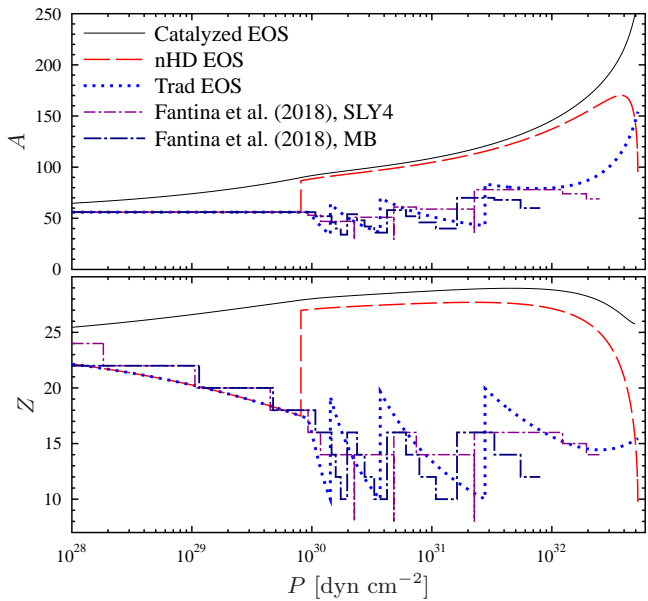}%
	\caption{
		The nuclei charge $Z$ and atomic mass number $A$ as function of pressure for different crustal EOSs.
		\label{Fig_Z}}
\end{figure}

The flat region at $\rho\sim 1.1\times 10^{12}$~g\,cm$^{-3}$ for Trad EOS 
corresponds to pycnonuclear reaction. 
These reactions are also clearly visible as jumps in Fig.\ \ref{Fig_Z}, 
which demonstrates profiles of nuclear charges $Z$ and mass numbers $A$ 
for the same EOSs as in Fig.\ \ref{Fig_EOS}
[in Supplementary material we also show the function $A_c(P)$].
In addition,
dot-dashed lines show profiles obtained in Ref.\ \cite{Fantina_ea18} 
ignoring the condition $\mu_n^\infty={\rm const}$.
The corresponding EOSs are calculated for Sly4 functional in the 
extended Thomas-Fermi approach
and for liquid-drop model of Ref.\ \cite{MB77}.
One can see that for traditional approach our CLDM  reproduces 
the results of Ref.\ \cite{Fantina_ea18} reasonably well, 
in particular, pycnonuclear reactions occur three times in the inner crust. 

Crust composition (i.e., $Z$ and $A$) for nHD EOS is determined by Eq.\ (\ref{diff}). 
One may note that it is remarkably different from that for Trad EOS, 
being rather close (at not too large $P$) 
to the composition of catalyzed crust.
The latter fact is not surprising, since at not too large $P$ 
two terms in the l.h.s.\ of Eq.\ (\ref{diff}) are much larger than the term $C \mu_n$ 
in its r.h.s., hence Eq.\ (\ref{diff}) is quite similar to its `catalyzed' counterpart, $\mu_N=0$. 
At larger $P$ surface tension decreases, because matter inside and outside nuclear clusters
becomes more and more similar, while the term $C \mu_n$ increases and, eventually, 
all three terms in Eq.\ (\ref{diff}) become comparable; 
as a result $Z$ and $A$ 
for nHD EOS substantially differ from those for catalyzed EOS at such $P$.

\textit{Discussion and conclusions. --}
%
We construct the model of the inner crust of accreting NS, 
which respects the nHD condition, $\mu_n^\infty={\rm const}$, 
imposed by the requirement of hydrostatic equilibrium with respect 
to superfluid equation in the most part of the inner crust 
and by the diffusion equilibrium in a thin layer near the outer-inner crust interface.
We find that the resulting nHD EOS is rather close to the catalyzed one, 
being significantly different from Trad EOS obtained in the traditional approach, 
which ignores the condition $\mu_n^\infty={\rm const}$
and
implicitly assumes that both nuclei and unbound neutrons move together with one and the same velocity.
Our another important result is that we found an instability that allows 
to transform nuclei into $npe$-matter at the crust-core boundary,
and explain its physical meaning.
We also demonstrate that the interface $P=P_{\rm oi}$ 
between the (accreted) outer and inner crust 
is not associated with the `standard' neutron drip pressure $P_{\rm nd}$, 
at which neutrons `drip out' of nuclei \cite{Chamel_etal15_Drip}.
Instead, $P_{\rm oi}$ 
is determined 
by the nHD equilibrium condition inside the star.
As a result, $P_{\rm oi}$ in the accreted crust 
appears to be just a bit higher
than in 
the catalyzed crust, 
and nuclei at the $P=P_{\rm oi}$ interface absorb neutrons rather than emit them,
as in Trad EOS.
Neutron absorptions (accompanied by electron emissions) 
lead to a jump of $A$ at the upper boundary of the inner crust (see Fig.\ \ref{Fig_Z}). 
Neutrons, necessary for such absorptions, are supplied by upward neutron flow,
which originates at the crust-core boundary, where nuclei disintegrate into neutrons
as a result of the instability discussed above. Then these neutrons redistribute
over the inner crust and core in order to maintain nHD equilibrium.

Similarity of nHD and catalyzed EOSs suggests that accretion should have 
a less pronounced effect
on the crust thickness and tidal deformability 
than in the traditional approach.
It also suggests that the heat release
due to nonequilibrium nuclear reactions in the accreted crust
should be  
much smaller than it is usually thought to be,
and this idea agrees with 
apparently very different reaction flows for nHD EOS
(e.g., pycnonuclear reactions for nHD EOS 
are absent).
The heat release problem is considered in detail in our forthcoming publication \cite{GC19_HeatReleaze}.
According to preliminary estimates, 
the net heat release is $\sim 0.5\div0.7$~MeV/nucleon 
(i.e., $2-3$ times smaller than in the traditional approach), 
with significant fractions released at the outer-inner crust interface and crust-core boundary.
These
findings,
along with modification of the 
transport properties and heat capacity (due to changed nuclear composition), 
should noticeably affect interpretation 
of transiently accreting NSs 
and may shed new light on the shallow heating and superburst ignition problems (e.g., \cite{mdkse18}).
The detailed analysis is left for future work.

The crucial role of the 
neutron hydrostatic and diffusion equilibrium for accreted crust EOS, revealed in this Letter, 
is a general feature, which can not be disregarded (see also \cite{CS19_NoEquil}).
However, we should warn the reader that our results are illustrated within the simplified CLDM, 
which treats nuclear mass and charge numbers as continuous variables 
and
neglects pairing and shell effects. 
In Ref.\ \cite{Fantina_ea18} the latter 
are shown to be important for the energy release in the traditional approach. 
According to our preliminary results, obtained within the nHD approach, 
shell effects mainly influence profile of the heat release and composition of the crust; 
at the same time, 
the total heat release
and $P(\rho)$ dependence are not strongly affected.

\begin{acknowledgments}
We are grateful to E.M.~Kantor and N.N.~Shchechilin for numerous useful discussions. 
Work is supported by Russian Science Foundation (grant 19-12-00133). 
\end{acknowledgments}

\newpage
\appendix
\begin{cbunit}
	\onecolumngrid
	\begin{center}
		{\bf \Large Supplementary material}
	\end{center}

\section{Compressible liquid-drop model}

In the Supplementary material we present more 
details about 
compressible liquid drop model (CLDM), 
used in the Letter.
For the reader's convenience, 
the text of the supplementary material is made self-contained.

\subsection{Energy density}
\label{Edensity}

Within the CLDM nuclei (clusters of nucleons) 
are described as a spherically symmetric liquid drops of nuclear matter, 
located at the center of spherical electrically neutral Wigner-Seitz (WS) cell \cite{lpr85,hpy07,ch08}.  
The model is parametrized by six parameters: 
$n_{ni}$, $n_{pi}$, and $n_{no}$ are the number densities of neutrons 
and protons inside, and neutrons outside nucleus, 
respectively 
(for simplicity, we assume that proton drip does not occur in the crust, 
which is
true for all numerical models discussed in the Letter); 
$\nu_{s}$ is the neutron surface density in the neutron skin;
$V_c$, $r_p$ are the WS cell volume 
and (proton) radius of a nucleus, respectively
(with such definition of the nucleus radius the proton skin vanishes, 
see, e.g., appendix B in Ref. \cite{lpr85} and Ref.\ \cite{DHM00}).

In addition, it is useful to introduce the volume fraction occupied by nucleus inside the WS cell, 
\begin{align}
w=\frac{4 \pi r_p^3}{3V_{c}};
\label{w}
\end{align}
the surface area of a nucleus, $\mathcal A=4\pi r_p^2$; 
and the electron number density, $n_e$, which is determined by the quasineutrality condition, $n_e=w n_{pi}$.
In what follows, we make use of the continuous CLDM, 
i.e., we treat the charge $Z=n_e V_c$ and mass number 
$A=Z+4\pi n_{ni} r_p^3/3+\mathcal A \nu_s$ of nuclei
as continuous variables, neglecting pairing and shell effects.
The energy density for CLDM is written as (see, e.g., Ref.\ \cite{hpy07})
\begin{eqnarray}
\epsilon&=&w\, \epsilon^\mathrm{bulk}(n_{ni},\, n_{pi})+(1-w)\,
\epsilon^\mathrm{bulk}(n_{no},\, 0)
+\frac{E_{s}(\nu_{s},r_{p})}{V_{c}}
+\frac{E_{C}(n_{pi},r_{p},w)}{V_{c}}
+\epsilon_e(n_e).
\label{edensity}
\end{eqnarray}
Here $\epsilon^\mathrm{bulk}(n_n,n_p)$ is the energy density 
as a function of neutron and proton number densities.
For numerical estimates we adopt SLY4 energy-density functional \cite{Chabanat_ea98_SLY4}; 
in the Letter the same functional is also used to calculate equation of state (EOS) in the neutron star (NS) core. 
In Eq.\ (\ref{edensity}) $E_C$ is the Coulomb energy,
\begin{equation}
E_C=\frac{16\,\pi^2}{15}(n_{pi} e)^2 r_p^5 f(w)
=\frac{3}{5}\frac{Z^2 e^2}{r_p}f(w), \label{EC}
\end{equation}
where $f(w)=1-1.5\,w^{1/3}+0.5\, w$; 
$e$ and $\epsilon_e$ is the electric charge and energy density of degenerate electron gas, 
respectively \cite{hpy07}. 
The electron gas is assumed to be ideal in all numerical calculations.

The surface energy is described, in a thermodynamically-consistent way, 
as (see, e.g., Refs.\ \cite{ll80,lpr85,hpy07})
\begin{equation}
E_s=\mathcal A (\sigma +\mu_{ns} \nu_s),
\end{equation}
where $\sigma$ and $\mu_{ns}$ are the so called surface tension and chemical potential 
for neutrons absorbed at the nucleus surface.
For simplicity, 
when calculating these quantities,	
we neglect curvature corrections, 
related to the fact that nucleus surface is a sphere rather than a plain.
Then $\sigma$ and $\mu_{ns}$ are the only functions of $\nu_s$: 
$\sigma=\sigma(\nu_s)$, $\mu_{ns}=\mu_{ns}(\nu_s)$ \cite{lpr85,hpy07}.
Introducing the total number of neutrons absorbed at the nucleus surface, 
$N_s=\mathcal{A} \nu_s$, the neutron chemical potential $\mu_{ns}$ can be presented as
\begin{equation}
\mu_{ns}=\left . \frac{\partial E_s}{\partial N_s}\right|_\mathcal A
=\frac{d}{d \nu_s}\left( \sigma+\mu_{ns}\nu_s\right).
\end{equation}
This expression implies a relation of thermodynamical consistency for the function $\sigma(\nu_s)$
\begin{equation}
\frac{d \sigma}{d \nu_{s}}=-\nu_s \frac{d \mu_{ns}}{d \nu_s}.
\label{ThermCons_sigma}
\end{equation}
For numerical estimates we extract
$\sigma(\nu_s)$ and $\mu_{ns}(\nu_s)$ from the results of Ref.\ \cite{DHM00}.
Because both $\sigma$ and $\mu_{ns}$ are functions of $\nu_s$,
one can also present $\sigma$ as a function of $\mu_{ns}$: $\sigma=\sigma(\mu_{ns})$ 
(see Fig.\ \ref{Fig_sigma}), which is especially suitable for numerical applications.

\begin{figure}
	\includegraphics[width=8cm]{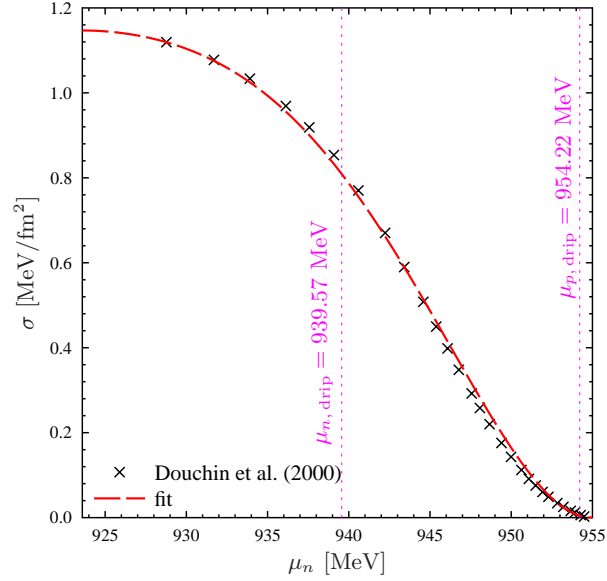}%
	\caption{Surface tension $\sigma$ versus $\mu_{ns}$. 
		Dashed line corresponds to the fit used in the Letter;
		crosses show the data extracted from Ref.\ \cite{DHM00}. 
		Thin vertical lines indicate neutron and proton drips.}
	\label{Fig_sigma}
\end{figure}

\subsection{Differential of the energy density and reduction to the two parameter EOS}
\label{LDA2}

To write down a differential of the energy density, $d \varepsilon$, in a compact form 
\cite{FootNoteApp},
%
%
let us present $\epsilon$ as function of  $n_{n{ i}}^{\rm (tot)}$, $n_{p{ i}}^{\rm (tot)}$, 
$n_{n{o}}^{\rm (tot)}$, $n_{s}^{\rm(tot)}$, $n_{N}$, and $w$:
$\epsilon=\epsilon(n_{n{ i}}^{\rm (tot)},\, n_{p{ i}}^{\rm (tot)},\, 
n_{n{ o}}^{\rm (tot)},\, n_{ns}^{\rm (tot)},\, n_{ N},\,w)$,
where the auxiliary variables are:
\begin{eqnarray}
n_{n{ i}}^{\rm (tot)} &\equiv& n_{n{i}} w,
\label{nnitot}\\
n_{p{i}}^{\rm (tot)} &\equiv& n_{p{ i}} w,
\label{npitot}\\
n_{n{ o}}^{\rm (tot)} &\equiv& n_{n{ o}} (1-w),
\label{nnotot}\\
n_{ns}^{\rm (tot)}&\equiv& \frac{N_{ s}}{V_{\rm cell}},
\label{nstot}
\\
n_N&=&V_c^{-1}.
\end{eqnarray}
They have a simple physical meaning.
For example, 
$n_{n{i}}^{\rm (tot)}$ is the number of neutrons in a nucleus per WS cell volume $V_c$, and
$n_{\rm N}$ is the number density of nuclei.
Using these variables, the baryon number density reads
\begin{equation}
n_b=n_{n{ i}}^{\rm (tot)}+n_{p{ i}}^{\rm (tot)}+n_{n{ o}}^{\rm (tot)}+n_{ns}^{\rm (tot)}, 
\label{nb}
\end{equation}
while the quasineutrality condition is simply $n_e=n_{p{ i}}^{\rm (tot)}$.
In terms of these variables $d \varepsilon$ can be written as
\begin{equation}
d \epsilon= \frac{\partial \epsilon }{\partial n_{n{ i}}^{\rm (tot)}} d n_{n{ i}}^{\rm (tot)}
+\frac{\partial \epsilon }{\partial n_{p{ i}}^{\rm (tot)}} d n_{p{ i}}^{\rm (tot)}
+\frac{\partial \epsilon }{\partial n_{n{ o}}^{\rm (tot)}} d n_{n{ o}}^{\rm (tot)}
+\frac{\partial \epsilon }{\partial n_{ns}^{\rm (tot)}} d n_{ns}^{\rm (tot)}
+\frac{\partial \epsilon }{\partial n_{ N}} d n_{ N}
+\frac{\partial \epsilon }{\partial w} dw,
\label{de2}
\end{equation}
or, after straightforward calculation of derivatives using Eq.\ (\ref{edensity}), as
\begin{eqnarray}
d \epsilon&=& \mu_{n{ i}} d n_{n{ i}}^{\rm (tot)}
+\left(\mu_{p{ i}}+\mu_e+\frac{2}{Z} E_C\right) d n_{p{ i}}^{\rm (tot)}
+ \mu_{n{ o}} d n_{n{ o}}^{\rm (tot)}
+\mu_{n{ s}} d n_{ns}^{\rm (tot)}
\nonumber\\
&+&\frac{1}{3}\left(\sigma \mathcal{A}-2  E_{\rm Coul}\right) d n_{ N}
+ \left[P_{ o}^\mathrm{bulk}-P_{ i}^\mathrm{bulk}+\frac{2 \sigma}{r_p}
-\frac{(1-w)\,n_N}{3\,w f(w)}\,E_C
\right] dw,
\label{de3}
\end{eqnarray}
where we defined
\begin{eqnarray}
\mu_{n{ i}}&=&
\mu_{n{ i}}^\mathrm{bulk}\equiv
\frac{\partial \epsilon^\mathrm{bulk}(n_{n{ i}},\, n_{p{ i}})}{\partial n_{n{ i}}},
\quad\quad
\mu_{n{ o}}=
\mu_{n{ o}}^\mathrm{bulk}\equiv
\frac{\partial \epsilon^\mathrm{bulk}(n_{n{ o}},\,0)}{\partial n_{n{ o}}},
\label{mun}\\
\mu_{p{ i}}&=&
\mu_{p{ i}}^\mathrm{bulk}\equiv 
\frac{\partial \epsilon^\mathrm{bulk}(n_{n{ i}},\, n_{p{ i}})}{\partial n_{p{ i}}},
\label{mupi}\\
\mu_e&=&\frac{\partial \epsilon_e(n_e)}{\partial n_e}\\
P_{ i}^\mathrm{bulk}&=&-\epsilon^\mathrm{bulk}(n_{n{ i}},\, n_{p{ i}})
+\mu_{p{ i}}^\mathrm{bulk}\, n_{p{ i}}
+\mu_{n{ i}}^\mathrm{bulk}\, n_{n{ i}},
\\
P_{ o}^\mathrm{bulk}&=&-\epsilon^\mathrm{bulk}(n_{n{ o}},\,0)+\mu^\mathrm{bulk}_{n{ o}}\, n_{n{ o}}.
\label{Pio}
\end{eqnarray}
Now, expressing $n_{n{ i}}^{({\rm tot})}$ through $n_b$ using Eq.\ (\ref{nb}),
Eq.\ (\ref{de3}) can be rewritten as
\begin{eqnarray}
d \epsilon&=& \mu_{n{ i}} d n_b
+\left(\frac{2 E_C}{Z}+\mu_{p{ i}}+\mu_e-\mu_{n{ i}}\right) d n_{p{ i}}^{\rm (tot)}
+ (\mu_{n{ o}}-\mu_{n{ i}}) d n_{n{ o}}^{\rm (tot)}
+(\mu_{n{ s}}-\mu_{n{ i}}) d n_{ns}^{\rm (tot)}
\nonumber\\
&+&\frac{1}{3}\left(\sigma \mathcal{A}-2  E_{\rm Coul}\right) d n_{ N}
+ \left[P_{ o}^\mathrm{bulk}-P_{ i}^\mathrm{bulk}+\frac{2 \sigma}{r_p}
-\frac{(1-w)\,n_N}{3\,w f(w)}\,E_C
\right] dw.
\label{de4}
\end{eqnarray}
This equation is ideally suited to minimize the energy density at fixed $n_b$ and $n_N$ with respect to other variables,
e.g., $n_{pi}^{\rm (tot)}$, $n_{no}^{\rm (tot)}$, $n_{ns}^{\rm (tot)}$, and $w$.
It leads to two-parameter EOS, $\epsilon(n_b, n_{N})$, discussed in the Letter. 
Other variables of CLDM, required to calculate $\varepsilon$ via Eq.\ (\ref{edensity}), 
can be determined, for given $n_b$ and $n_N$, from the following equations 
[see Eq. (\ref{de4})]:\\
1) Beta-equilibrium condition (minimization of $\epsilon$ with respect to $n_{pi}^\mathrm{(tot)}$):
\begin{equation}
\mu_{n{ i}}=\mu_{p{ i}}+\mu_e+\frac{2 E_C}{Z}; \label{Beta-Equl}
\end{equation}
2) Neutron local diffusion equilibrium 
inside the WS cell 
(minimization of $\epsilon$ with respect to $n_{no}^\mathrm{(tot)}$ and $n_{ns}^\mathrm{(tot)}$ )
\begin{equation}
\mu_{n{ i}}=\mu_{no}=\mu_{ns}; \label{mun_equl}
\end{equation}
3) The pressure balance equation (minimization of $\epsilon$ with respect to $w$)

\begin{equation}
P_{ i}^\mathrm{bulk}=P_{ o}^\mathrm{bulk}+\frac{2 \sigma}{r_p}
-\frac{(1-w)\,n_N}{3\,w f(w)}\,E_C.
\label{Press-bal}
\end{equation}

According to Eq.\ (\ref{mun_equl}), 
the neutron chemical potential is constant in all parts of a WS cell 
(inside and outside of a nucleus, as well as in the neutron skin), 
as it should be in thermodynamic equilibrium.
Thus, when considering the two-parameter EOS, $\epsilon(n_b, n_{N})$,
it is reasonable to suppress unnecessary indices and  
denote the neutron chemical potential simply as $\mu_n$.
Using Eqs.\ (\ref{Beta-Equl})--(\ref{Press-bal}), 
the expression (\ref{de4}) reduces to
\begin{equation}
d \epsilon= \mu_{n} d n_b+\mu_N d n_{ N}, \label{de_twoparams}
\end{equation}
where we introduced a new parameter,
\begin{equation}
\mu_N=\frac{1}{3} (\sigma \mathcal A- 2 E_C), \label{mu_N}
\end{equation}
which can be interpreted as an effective chemical potential for nuclear clusters.
It describes how the system energy will change if we create there an additional nuclear cluster, 
keeping, at the same time, 
the total baryon number in the system fixed.

\subsection{Pressure}

To finalize construction of EOS we should determine the pressure $P$. 
According to the second law of thermodynamics, it is equal to (minus) derivative of the system energy with respect to volume $V$ at fixed total particle numbers
$n_{n{ b}} V$, $n_{p{ i}}^{\rm (tot)} V$, $n_{n{ o}}^{\rm (tot)} V$,
$n_{ns}^{({\rm tot})}V$, and $n_{ N} V$.
Thus, in the most general case, using Eq.\ (\ref{de4}), 
the pressure can be written as
\begin{equation}
P\equiv-\frac{\partial (\epsilon V)}{\partial V}=-\epsilon + \mu_{n{ i}}n_b
+\left(\frac{2 E_C}{Z}+\mu_{p{ i}}+\mu_e-\mu_{n{ i}}\right)\,n_{p{ i}}^{\rm (tot)}+(\mu_{n{ o}}- \mu_{n{ i}})n_{n{ o}}^{\rm (tot)}+(\mu_{n{ s}}- \mu_{n{ i}}) n_{ns}^{\rm (tot)}
+\mu_N n_N.
\label{P3}
\end{equation}
For a two-parameter EOS, $\epsilon(n_b, n_{N})$, determined by the minimization procedure discussed 
above, 
$P$ reduces to [see Eqs.\ (\ref{Beta-Equl})--(\ref{Press-bal})]
\begin{equation}
P=-\epsilon + \mu_{n}n_b+\mu_N n_{ N}.
\label{P4}
\end{equation}
%

\section{Absence of solutions at $P>P_{\rm max}$ as a generic feature of equation (3) from the Letter}
\label{instability1}

Let us start with equation (3) from the Letter 
that should be used to determine $n_{\rm N}$ for a given $n_b$
for neutron Hydrostatic and Diffusion (nHD) EOS,
\begin{align}
\sigma \mathcal{A}-2  E_{C}=3 C \mu_n,
\label{vir1}
\end{align}
In fact, in what follows it will be more convenient to solve it for the nucleus radius $r_p$ 
(rather than $n_{N}$),
and then find $n_{N}$ from the formula (\ref{w}) and the relation $n_{N}=1/V_c$.
Using the expression (\ref{EC}) for $E_C$, Eq.\ (\ref{vir1}) can be rewritten as
\begin{equation}
\sigma(\mu_n) = \frac{8\pi}{15} \, n_{pi}^2 e^2 f(w) \, r_p^3 + \frac{3\, C \mu_n}{4\pi r_p^2},
\label{vir2}
\end{equation}
where 
the surface tension $\sigma$ 
is presented as the (known) function 
of the neutron chemical potential, $\mu_n=\mu_{ns}$ 
(see Fig.\ \ref{Fig_sigma} and a comment at the end of Sec.\ \ref{Edensity}).

The quantities $w$, $n_{pi}$, and $\mu_n$ in Eq.\ (\ref{vir2}) 
can be thought of as already expressed, using Eqs.\ (\ref{Beta-Equl})--(\ref{Press-bal}),
as some functions of $r_p$ and $n_b$.
Thus, Eq.\ (\ref{vir2}) is, generally, an implicit equation for $r_p$, which is difficult to 
solve and analyze.
However, the analysis can be substantially simplified if one notes 
that such quantities as $n_{ni}$, $n_{pi}$, $n_{no}$, $w$, $\mu_n$
can be found, to a good precision, from the so called 
bulk approximation, which ignores surface and Coulomb effects (see, e.g., Ref.\ \cite{DHM00}).
In other words, these quantities should be approximately the same, 
e.g., for catalyzed matter and for nHD EOS we are interested in 
(these EOSs do not differ on the level of bulk approximation).
Therefore, 
for a qualitative analysis of Eq.\ (\ref{vir2}) 
we can present
$w$ and $n_{pi}$ in Eq.\ (\ref{vir2}) 
as the same functions of $\mu_n$ as for catalyzed matter.
Since in this approximation $w$ and $n_{pi}$ do not depend on $r_p$,
Eq.\ (\ref{vir2}) can now be readily solved for $r_p$.
However, we do not really need this solution in order to demonstrate 
that it does not have roots at some $P>P_{\rm max}$, or, equivalently, at $\mu_n>\mu_{n \, {\rm max}}$
(because $\mu_n$ is a growing function of $P$
for nHD EOS).
To do that, note that the surface tension $\sigma(\mu_n)$
[the function in the left-hand side of Eq.\ (\ref{vir2})] 
is a decreasing function of $\mu_n$ (see Fig.\ \ref{Fig_sigma}),
since nuclear matter inside and outside nuclear clusters 
becomes more and more similar with growing density \cite{DHM00}. 
At the same time, the right-hand side of Eq.\ (\ref{vir2})
has a minimum as a function of $r_p$ at a point
\begin{align}
r_{p0}=\left( \frac{15  C \mu_n}{16\pi^2 n_{pi}^2 e^2 f(w)}\right)^{1/5}.
\label{rpmin}
\end{align}
The corresponding minimum value of the right-hand side 
[let us denote it $F(\mu_n)$],
which is a function of $\mu_n$, is given by
\begin{equation}
F(\mu_n)=\left(\frac{125\, C^3}{36\,\pi}\,n_{pi}^4 e^4 f^2(w) \mu_n^3 \right)^{1/5},
\label{min}
\end{equation}
and is a growing function of $\mu_n$.
Correspondingly, because the left-hand side of Eq.\ (\ref{vir2}) decreases, 
while the right-hand side increases with $\mu_n$, 
the maximum reachable value of the neutron chemical potential 
in the crust, $\mu_n=\mu_{n\, {\rm max}}$, 
is a solution to the equation
\begin{align}
\sigma(\mu_n)=F(\mu_n).
\label{vir3}
\end{align}
At $\mu_n>\mu_{n\,{\rm max}}$ the solution does not exist and we have an instability 
discussed in the Letter.
Note that $n_{pi}$ and $w$ in this equation are (known) functions of $\mu_n$.
Numerical solution to Eq.\ (\ref{vir3}) is shown by 
dashes
in Fig.\ \ref{figsupppl2}, 
where we also present the function $\mu_{n\, {\rm max}}(C)$ 
obtained
directly from equation (\ref{vir1}) [or (\ref{vir2})]
without any approximations 
(solid line).
As one may see the agreement between the curves is rather good, 
which additionally justifies the approach described above.
\begin{figure}
	\includegraphics[width=8cm]{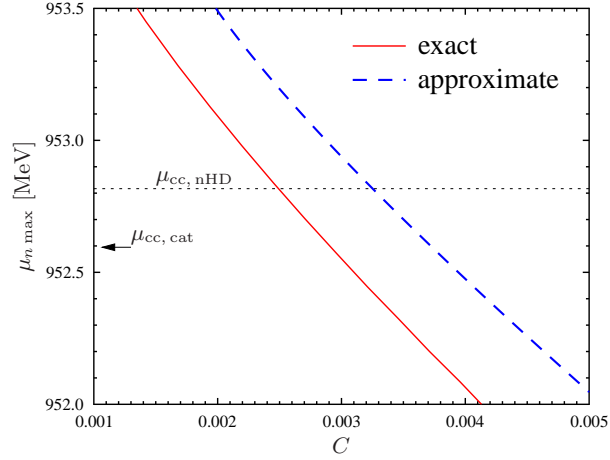}%
	\caption{Maximum reachable neutron chemical potential
		in the crust, $\mu_{n \, {\rm max}}$, 
		as a function of the parameter $C$. 
		Dashes: $\mu_{\rm n \, {\rm max}}(C)$ is obtained using 
		the approximate method described in the text. 
		Solid line: $\mu_{\rm n \, {\rm max}}(C)$
		is numerically calculated from Eq.\ (\ref{vir1}) without any approximations.
		Dotted horizontal line and arrow show $\mu_n$
		at the crust-core boundary for  
		the fully accreted nHD EOS ($\mu_n=\mu_{{\rm cc},\, {\rm nHD}}$)
		and for catalyzed crust ($\mu_n=\mu_{{\rm cc},\, {\rm cat}}$), respectively.
	}
	\label{figsupppl2}
\end{figure}

\section{Crustal composition for different EOSs}

In Fig. \ref{Fig_Z} we present profiles of the crustal composition 
for several EOSs discussed in 
the Letter. 
The figure is analogue of figure 2 from the main text, 
with additional panel showing the total number of nucleons in a WS cell, $A_c$.

\begin{figure}
	\includegraphics[width=8cm]{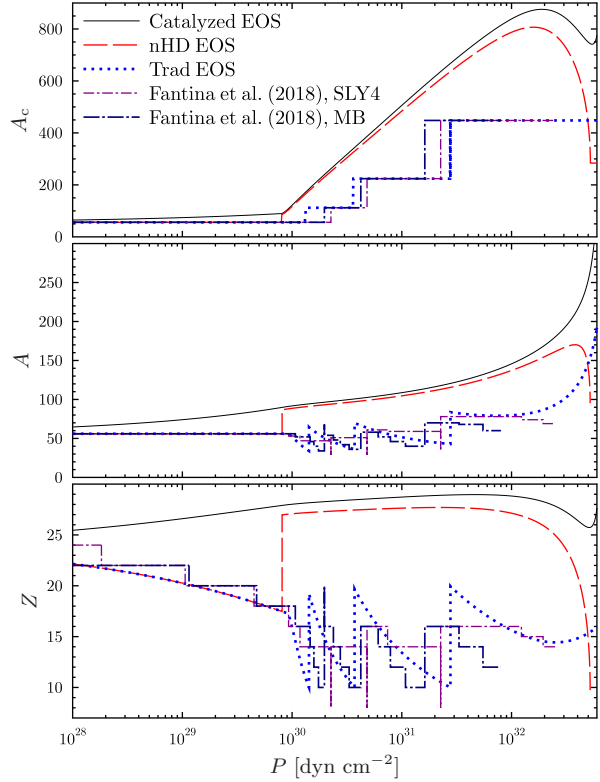}%
	\caption{The nuclei charge $Z$, atomic mass number $A$ and total number of nucleons in a WS cell, $A_c$, as function of pressure for different crustal EOSs discussed in the Letter.\label{Fig_Z}}
\end{figure}



\begin{thebibliography}{6}
\bibitem{FootNoteApp}
	Using Eq.\ (\ref{edensity}), it is straightforward to present $d \varepsilon$ 
		in terms of original variables ($n_{ni}$, $n_{pi}$, $n_{no}$, $w$, $V_C$, $\nu_s$), 
		but the resulting equations will be rather lengthly.
\end{thebibliography}

%

\end{cbunit}

\end{document}